\documentclass[useAMS,usenatbib]{aa}
\usepackage{txfonts,graphicx,amssymb,subfigure}
\usepackage[normalem]{ulem}

\usepackage{natbib}

\begin{document}

\title{Recognizing magnetic structures by present and future radio
telescopes with Faraday rotation measure synthesis}

\author{R. Beck\inst{1}\fnmsep
\thanks{Corresponding author: \email{rbeck@mpifr-bonn.mpg.de}}
P. Frick\inst{2}\inst{,3}, R. Stepanov\inst{2}\inst{,3}, and D.
Sokoloff\inst{4}}
\titlerunning{Recognizing magnetic structures with RM synthesis}
\authorrunning{R. Beck et al.}
\institute{MPI f\"ur Radioastronomie, Auf dem H\"ugel 69, 53121
Bonn, Germany \and Institute of Continuous Media Mechanics, Korolyov
str. 1, 614013 Perm,  Russia \and State National Research
Polytechnical University of Perm, Komsomolskii av. 29, 614990 Perm,
Russia \and Department of Physics, Moscow State University, 119991
Moscow, Russia}

\date{Received 22 February 2012; accepted 29 April 2012}

\abstract{Modern radio telescopes allow us to record a large number
of spectral channels. The application of a Fourier transform to
spectropolarimetric data in radio continuum, Faraday rotation
measure (RM) synthesis, yields the ``Faraday spectrum'', which hosts
valuable information about the magneto-ionic medium along the line
of sight.} {We investigate whether the method of wavelet-based RM
synthesis can help us to identify  structures of regular and
turbulent magnetic fields in extended magnetized objects, such as
galaxies and galaxy clusters.} {The analysis of spectropolarimetric
radio observations of multi-scale targets calls for a corresponding
mathematical technique. Wavelets allow us to reformulate the RM
synthesis method in a scale-dependent way and to visualize the data
as a function of Faraday depth and scale.} {We present observational
tests to recognize magnetic field structures. A region with a
regular magnetic field generates a broad ``disk'' in Faraday space,
with two ``horns'' when the distribution of cosmic-ray electrons is
broader than that of the thermal electrons. Each field reversal
generates one asymmetric ``horn'' on top of the ``disk''. A region
with a turbulent field can be recognized as a ``Faraday forest'' of
many components. These tests are applied to the spectral ranges of
various synthesis radio telescopes. We argue that the ratio of
maximum to minimum wavelengths determines the range of scales that
can be identified in Faraday space.} {A reliable recognition of
magnetic field structures in spiral galaxies or galaxy clusters
requires the analysis of data cubes in position--position--Faraday
depth space (``PPF cubes''), observed over a wide and continuous
frequency range, allowing the recognition of a wide range of scales
as well as high resolution in Faraday space. The planned Square
Kilometre Array (SKA) will fulfill this condition and will be close
to representing a perfect ``Faraday telescope''. The combination of
data from the Low Frequency Array (LOFAR, at low frequencies) and
the Expanded Very Large Array (EVLA, at high frequencies) appears to
be a promising approach for the recognition of magnetic structures
on all scales. The addition of data at intermediate frequencies from
the Westerbork Synthesis Radio Telescope (WSRT) or the Giant
Meterwave Radio Telescope (GMRT) would fill the gap between the
LOFAR and EVLA frequency ranges. The Global Magneto-Ionic Medium
Survey (GMIMS), planned with several single-dish telescopes at low
angular resolution, will also provide good scale recognition and
high resolution in Faraday space.}


\maketitle


\section{Introduction}

Modern radio telescopes are equipped with digital correlators that
allow us to record a large number of spectral channels. While radio
spectroscopy in total intensity is well-developed, the possibilities
of spectropolarimetry in radio continuum have been explored for only
a few years. The fundamentals were presented by \cite{burn66}, while
the first application to multi-channel polarization data (data
cubes) was presented by \cite{brentjens05}.

Faraday rotation measure (RM) synthesis generates the ``Faraday
dispersion function'' or, in short, the ``Faraday spectrum''
$F(\phi)$, which is the (complex-valued) polarized intensity
spectrum as a function of ``Faraday depth'' $\phi$
\begin{equation}
\phi(x) = 0.81\int_{0}^{x} B_\parallel (x') n_{\rm e}(x')
\mathrm{d}x', \label{fardep}
\end{equation}
where $B_\parallel$ is the line-of-sight magnetic field component
measured in $\mu$G, $n_{\rm e}$ is the thermal electron density
measured in cm$^{-3}$, and the integral is taken along the line of
sight through the region containing both magnetic fields and thermal
electrons, where $x'$ is measured in parsecs. Our adopted convention
is that $\phi$ is positive when $\mathbf{B}$ is pointing towards the
observer.

\cite{burn66} showed that the complex polarized intensity $P$ at
different wavelengths $\lambda$ can be calculated from $F(\phi)$ via
a Fourier transform
\begin{equation}
\label{p_to_f} P(\lambda^2) =  \int_{-\infty}^{\infty} F(\phi) {\rm
e}^{2{\rm i}\phi \lambda^2}  \mathrm{d} \phi ,
\end{equation}
which means that the Faraday spectrum $F$ is the Fourier transform
of the complex polarized intensity:
\begin{equation}
\label{f_to_p1} F(\phi) = {{1} \over{\pi}}\int_{-\infty}^{\infty}
P(\lambda^2) {\rm e}^{-2{\rm i}\phi \lambda^2}  \mathrm{d} \lambda^2
. \label{Burn}
\end{equation}
Equation (\ref{Burn}) is the basis of RM synthesis. A major
limitation emerges from the fact that $P$ can be measured only for
$\lambda^2>0$ and practically only in a finite spectral band
$\lambda_{\rm min} < \lambda < \lambda_{\rm max}$.

The quantity $F$ can be used to determine the Faraday depth and the
intrinsic polarization angle of each component in the Faraday
spectrum. As in classical spectroscopy, the interpretation of this
spectrum is not straightforward. In particular, there is no simple
relation between Faraday depth and geometrical depth. Furthermore,
the Faraday spectrum suffers from sidelobes of the main components
caused by limited coverage of the wavelength space, and in the case
of point-like ``sources'' these sidelobes can be removed by ``RM
clean'' \citep{heald09}.

RM synthesis is characterized by three basic parameters
\citep{brentjens05}:
\begin{itemize}
\item{the resolution $\delta\phi$ in Faraday space,
which is inversely proportional to the coverage $\Delta\lambda^2$ in
wavelength ($\lambda^2$) space;}
\item{the maximum observable $|\phi_{\rm max}|$ of
point-like sources in Faraday space, which is inversely proportional
to the width of a single frequency channel;}
\item{the maximum width $|\Delta\phi_{\rm max}|$ of extended
structures in Faraday space (Faraday-rotating {\em and}\
synchrotron-emitting regions), which is inversely proportional to
the square of the minimum observation wavelength. Wide-band
observations at long wavelengths yield high resolution in Faraday
space but cannot detect extended structures.}
\end{itemize}
We argue below (Sect.~3) that the ratio of maximum to minimum
wavelengths is another important parameter in helping us to
recognize a range of scales in Faraday space.

\begin{table}
\caption{Spectral ranges of various radio telescopes and parameters
crucial for RM synthesis (see text for details).} \label{Tab1}
 \centering
   \begin{tabular}{|l|c|c|c|c|c|}
   \hline
{\bf Telescope} & $\lambda$ & {\bf $\Delta \lambda^2$}
& $|\delta\phi|$ & $|\Delta\phi_{\rm max}|$ & $(\lambda_{\rm max}/$ \\
& m & m$^2$ & rad/m$^2$ & rad/m$^2$ & $\lambda_{\rm min})^2$\\
   \hline
LOFAR HBA & 1.25--2.73 & 5.9 & 0.59 & 2.8 & 4.8 \\
   \hline
WSRT & 0.17--0.23 && && \\
& + 0.77--0.97 & 0.91$^1$ & 3.8 & 110 & 33 \\
   \hline
GMRT & 0.21--0.30 && &&\\
& + 0.47--0.52 && && \\
& + 0.87--0.98 & 0.92$^1$ & 3.8 & 71 & 22 \\
   \hline
DRAO, Parkes, & 0.17--0.23 && &&\\
Effelsberg & + 0.33--1.0 & 0.97 & 3.6 & 110 & 35 \\
(GMIMS) &&&&&\\
   \hline
Parkes (S-PASS) & 0.12--0.14 & 0.004 & 870 & 220 & 1.4 \\
   \hline
Arecibo & 0.20--0.24 & 0.021 & 165 & 79 & 1.4 \\
(GALFACTS) &&&&&\\
   \hline
EVLA & 0.025--0.30 & 0.089 & 39 & 5000 & 144 \\
   \hline
ATCA & 0.03--0.27 & 0.072 & 48 & 3500 & 81 \\
   \hline
ASKAP & 0.21--0.27 & 0.026 & 130 & 71 & 1.6 \\
(POSSUM) &&&&&\\
(POSSUM & 0.21--0.42 & 0.14 & 25 & 71 & 4.0 \\
+ FLASH) &&&&&\\
   \hline
SKA phase 1 & 0.10--4.3 & 18 & 0.19 & 310 & 1850 \\
SKA phase 2 & 0.03--4.3 & 18 & 0.19 & 3500 & 20500 \\
   \hline
\end{tabular}
\newline $^1$ High sidelobes in Faraday spectrum expected owing to the large gaps in wavelength coverage
\end{table}


\begin{figure}
\centering
\includegraphics[width=0.45\textwidth]{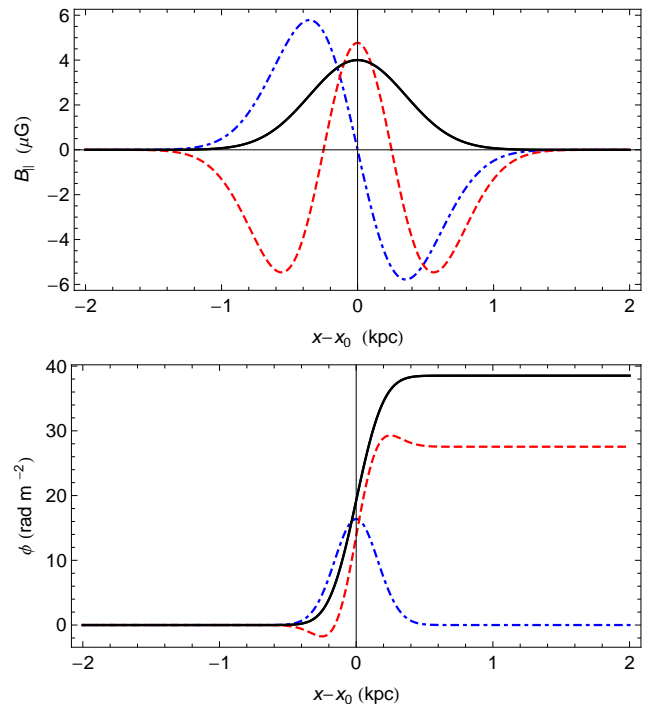}
\caption{Several typical examples of a region with and without
reversals of the regular magnetic field: the solid black line shows
the Gaussian magnetic field distribution, the dot-dashed blue line
stands for one reversal and the dashed red line shows the
distribution with two reversals. {\bf Top:} Magnetic field profile
$B(x)$ along the line of sight. {\bf Bottom:} Faraday depth
distribution $\phi(x)$ along the line of sight.} \label{Bfig}
\end{figure}


\begin{figure}
\centering
\includegraphics[width=0.45\textwidth]{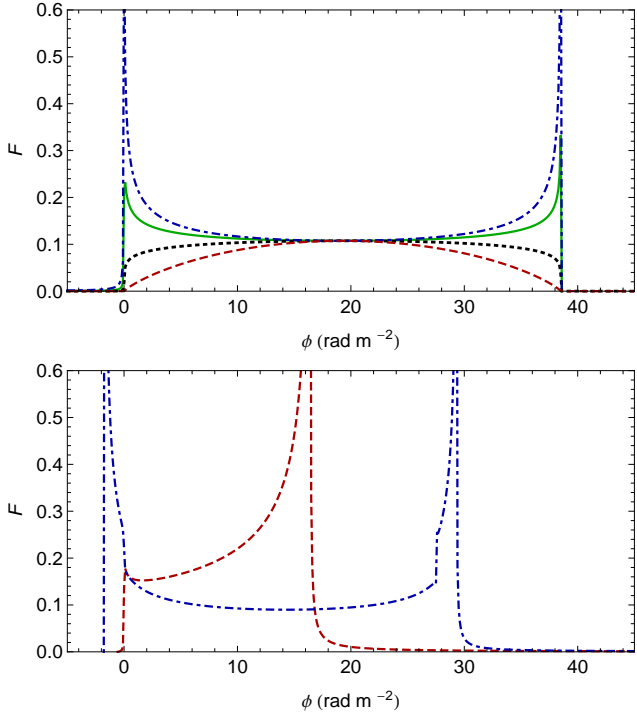}
\caption{Several typical examples of the Faraday spectrum in Faraday
depth space $F(\phi)$. {\bf Top:} Gaussian magnetic field
distribution for four different models (see Table~2): solid green
line (model 1); dashed black line (model 2); dashed red line (model
3); dot-dashed blue line (model 4). {\bf Bottom:} The case of
magnetic field reversals and model 1): one reversal (dot-dashed blue
line); two reversals (dashed red line).} \label{Ffig}
\end{figure}

\begin{table}
 \centering
   \begin{tabular}{|l|c|c|c|}
   \hline
Model & $h_{\rm c}/h_{\rm th}$ & $h_{\rm c}/h_{\rm B}$ &
Remark\\
   \hline
1 & $\sqrt 2$ & $1/\sqrt 2$ & Equipartition between $n_{\rm c}$ and $B$\\
2 & $1$ & $1/2$ &  Similar distributions of $n_{\rm c}$ and $n_{\rm th}$\\
3 & $1/\sqrt 2$ & $1/(2\sqrt 2)$ & $n_{\rm th}$ more extended \\
4 & $2$ & $1$ & $n_{\rm c}$ more extended \\
   \hline
\end{tabular}
\caption{Models of the density distributions of cosmic-ray electrons
($n_{\rm c}$) and thermal electrons ($n_{\rm th}$) considered in
this paper. $h_{\rm c}$ is the Gaussian scale-height (or
scale-radius) of the distribution of cosmic-ray electrons, $h_{\rm
c}$ that of the thermal electrons, and $h_{\rm B}$ that of the
regular magnetic field $B$ (see Section~4 for details).}
\label{Tab2}
\end{table}

A region of any extent in physical space that is
synchrotron-emitting but not Faraday-rotating (i.e. no thermal
electrons) generates a ``point source'' in Faraday space at
$\phi=0$. A non-emitting region with thermal electrons and magnetic
fields in front of the emitting region is called ``Faraday screen'';
it shifts the point source in Faraday space to a non-zero value of
$\phi$.

RM synthesis has been applied to spectropolarimetric data from the
Westerbork telescope \citep{bruyn05,schnitzeler09,heald+09,mao10,
brentjens11,pizzo11}, from the Parkes telescope and the Australia
Telescope Compact Array (ATCA) \citep{feain09,harvey10,sullivan12},
from the VLA \citep{vaneck11} and from the EVLA \citep{heesen11}.
The $\Delta\lambda^2$ coverages of these observations were small, so
that the resolution in Faraday space was limited. Nevertheless, many
Faraday spectra have revealed complex structures
\citep{brentjens11}. By means of the application of RM synthesis,
the fine-scale structure of the magnetic field in the Milky Way
around a local HI bubble \citep{wolleben10a} and around a supernova
remnant \citep{harvey10} could be measured.

On the other hand, as RM synthesis is regularly applied in radio
astronomy, the number of questions discussed about its technical
problems (algorithms, software) and practical limitations for
existing instruments similarly increases
\citep{heald09,farnsworth11,frick11,li11,andrecut11,bell11a}. A
combination of RM synthesis and two-dimensional image synthesis into
three-dimensional (3-D) ``Faraday synthesis'' of data from synthesis
telescopes has been proposed as a possible way of improving the
technique \citep{bell12}.

In this paper, we reconsider the problem of observations of large
magnetized objects, such as galaxies or the intergalactic medium in
galaxy clusters. Two factors are crucial in this context: the
objects are very extended and composed of numerous structures of
very different spatial scales, namely galaxies containing a disk,
halo, spiral arms, turbulent star-forming regions, and supernova
remnants, and galaxy clusters containing radio galaxies, jets,
outflows, turbulent gas, and ``relic'' shock fronts. The multi-scale
structure calls for the use of a mathematical technique developed in
the framework of wavelet analysis. Wavelets do not only provide a
more robust algorithm on which to base the RM synthesis method
\citep{frick10}, but can also be used as an illustrative tool for
the presentation of results. We will determine here the information
that can be recovered from RM synthesis based on wide-band
observations with existing and future radio telescopes.

For simplicity, we neglect in this paper the effects of different
angular resolutions and, in the case of synthesis telescopes, of
different distributions of baselines (uv coverage) when combining
data from different telescopes. We also neglect the spectral
variations in the polarized emission and the instrumental noise.


\begin{figure}
\centering
\includegraphics[width=0.45\textwidth]{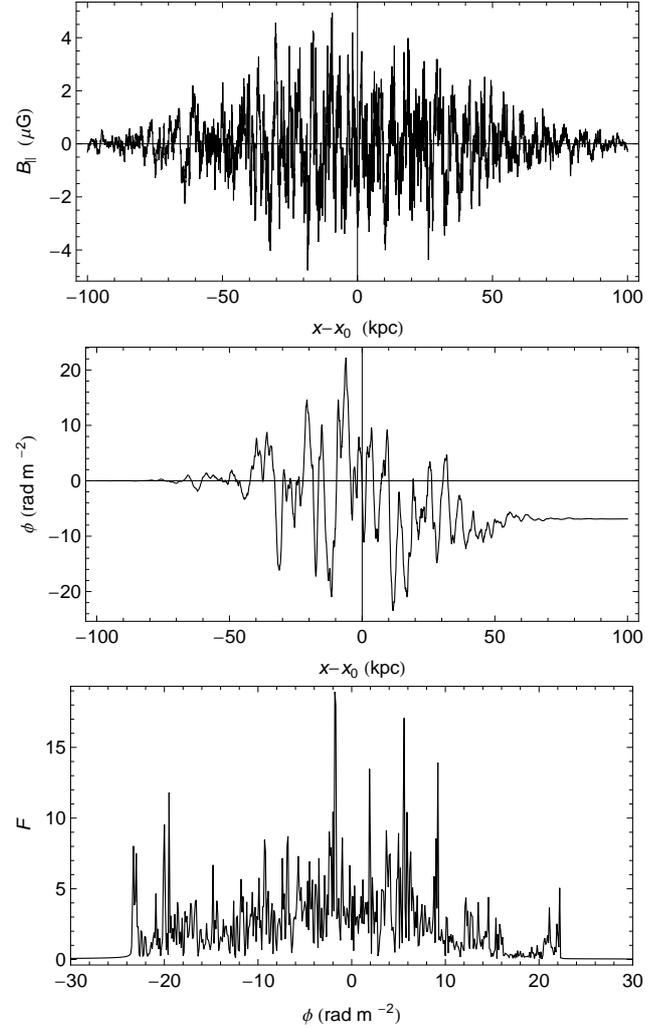}
\caption{Model example of a turbulent magnetic field in the
intergalactic medium of a galaxy cluster. Panels from above:
magnetic field profile along the line of sight $B(x)$, Faraday depth
distribution along the line of sight $\phi(x)$, and Faraday spectrum
in Faraday depth space $F(\phi)$. } \label{Btfig}
\end{figure}

\section{Present and future radio telescopes}

Spectropolarimetric data can be obtained with the following present
and future radio telescopes:

- Low Frequency Array (LOFAR, Europe), which is the first
online-connected synthesis telescope at low frequencies and operates
since 2011 \citep{stappers11}. Polarization from pulsars has been
detected in the ``lowband'' (10--80\,MHz) and in the lower
``highband'' (110--190\,MHz). Observations in the lowband, lower
highband, and higher highband (about 170--240\,MHz) need different
setups and cannot be performed simultaneously. The lower highband
has maximum sensitivity and is used to search for diffuse polarized
emission, which is one task of the ``magnetism key science project''
\citep{beck10}. The lowband has lower sensitivity and signals suffer
from strong Faraday depolarization. In this paper, the full
frequency range of the ``highband'' (HBA) of 110--240\,MHz (or about
1.25--2.7\,m) is considered.

- Westerbork Synthesis Radio Telescope (WSRT, the Netherlands),
which has successfully supplied spectropolarimetric data in the
frequency range 1300--1763\,MHz of a sample of nearby galaxies, the
SINGS survey \citep{heald+09}, clusters \citep{pizzo11}, and to
measure RMs of background sources \citep{mao10}. The low resolution
in Faraday space only allowed measurement of one $\phi$ component at
most locations, except for the central regions of several galaxies
where three components were detected. The lower frequency range
310--390\,MHz (or 0.77--0.97\,m) allows higher resolution in Faraday
space and hence is of great interest to RM synthesis, e.g. as
applied to data of the diffuse Galactic foreground
\citep{schnitzeler09} and of galaxy clusters
\citep{bruyn05,brentjens11,pizzo11}. A survey of nearby galaxies is
underway in this frequency band. After installation of the APERTIF
system, the WSRT will only observe around 1\,GHz.

- Giant Meterwave Radio Telescope (GMRT, India), a synthesis
telescope that operates in the frequency ranges 150--156, 236--244,
305--345, 580--640, and 1000--1450\,MHz. The polarization
calibration of the GMRT is more difficult than that of the WSRT, and
only a few successful detections of polarized signals from bright
radio sources in the two highest frequency bands have been reported
so far \citep{joshi10}. In the following we assume that the
frequency range 305--1450\,MHz can be used for future polarization
observations, which is similar to that of the WSRT.

- DRAO 26\,m single dish (Penticton, Canada), for which a northern
sky polarization survey (GMIMS) is ongoing in the frequency range
1277--1762\,MHz \citep{wolleben10b}.

\begin{figure*}
\begin{picture}(240,500)(0,0)
\put(0,000){\includegraphics[width=0.9\textwidth]{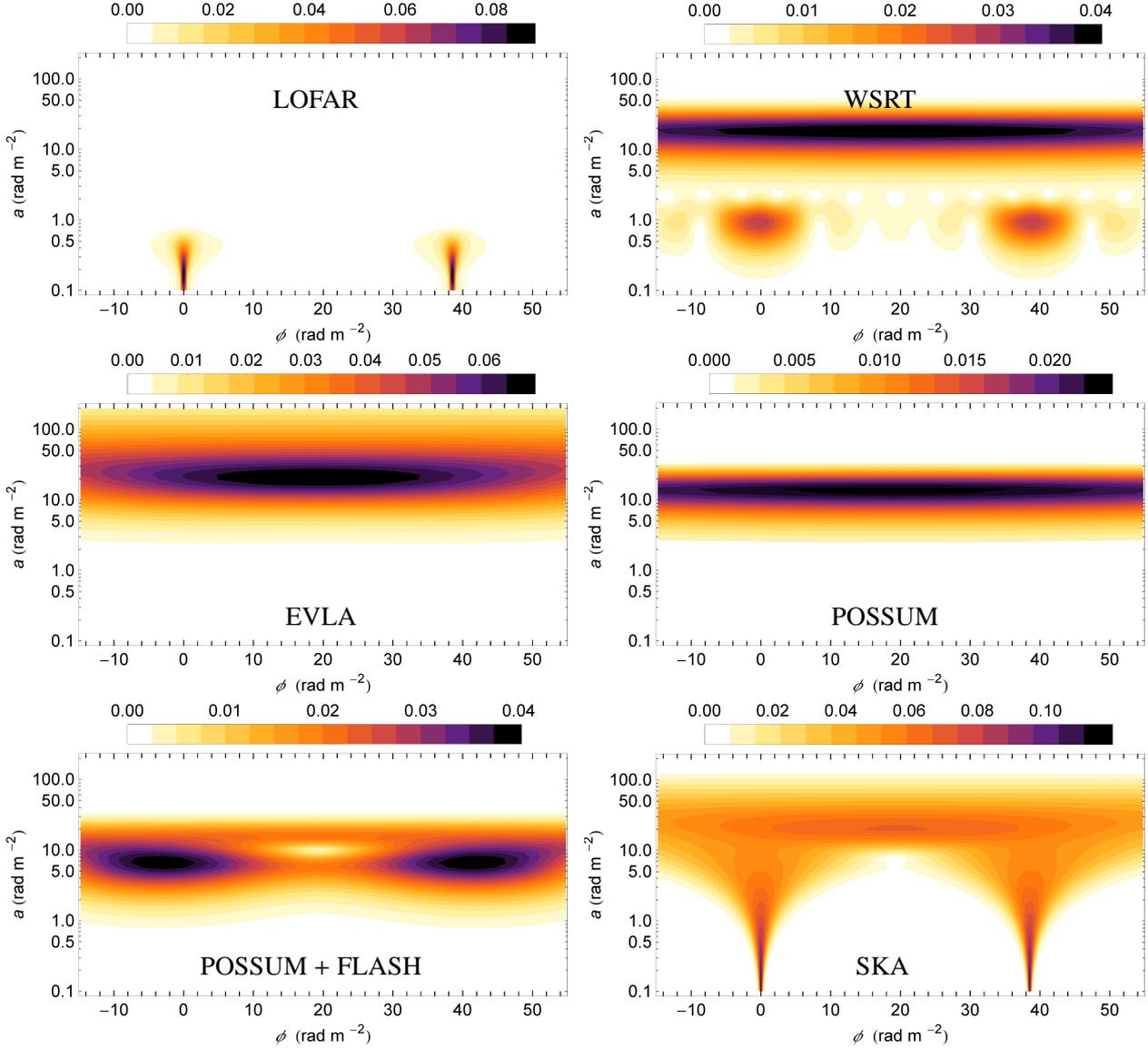}}
\put(110,390){{\large LOFAR}} \put(345,390){{\large WSRT}}
\put(115,175){{\large EVLA}} \put(340,175){{\large POSSUM}}
\put(80,30){{\large POSSUM + FLASH}} \put(350,30){{\large SKA}}
\end{picture}
\caption{Wavelet planes $w_F(a,\phi)$ for the magnetic field with a
Gaussian profile and in energy equipartition with cosmic rays (model
1). The calculations are done for the frequency bands available for
several telescopes. Left column, panels from above: LOFAR, EVLA (or
ATCA), and ASKAP (POSSUM + FLASH); right column, panels from above:
WSRT (or GMRT), ASKAP (POSSUM), and SKA phase 1. } \label{E1fig}
\end{figure*}



\begin{figure*}
\begin{picture}(240,500)(0,0)
\put(0,000){\includegraphics[width=0.9\textwidth]{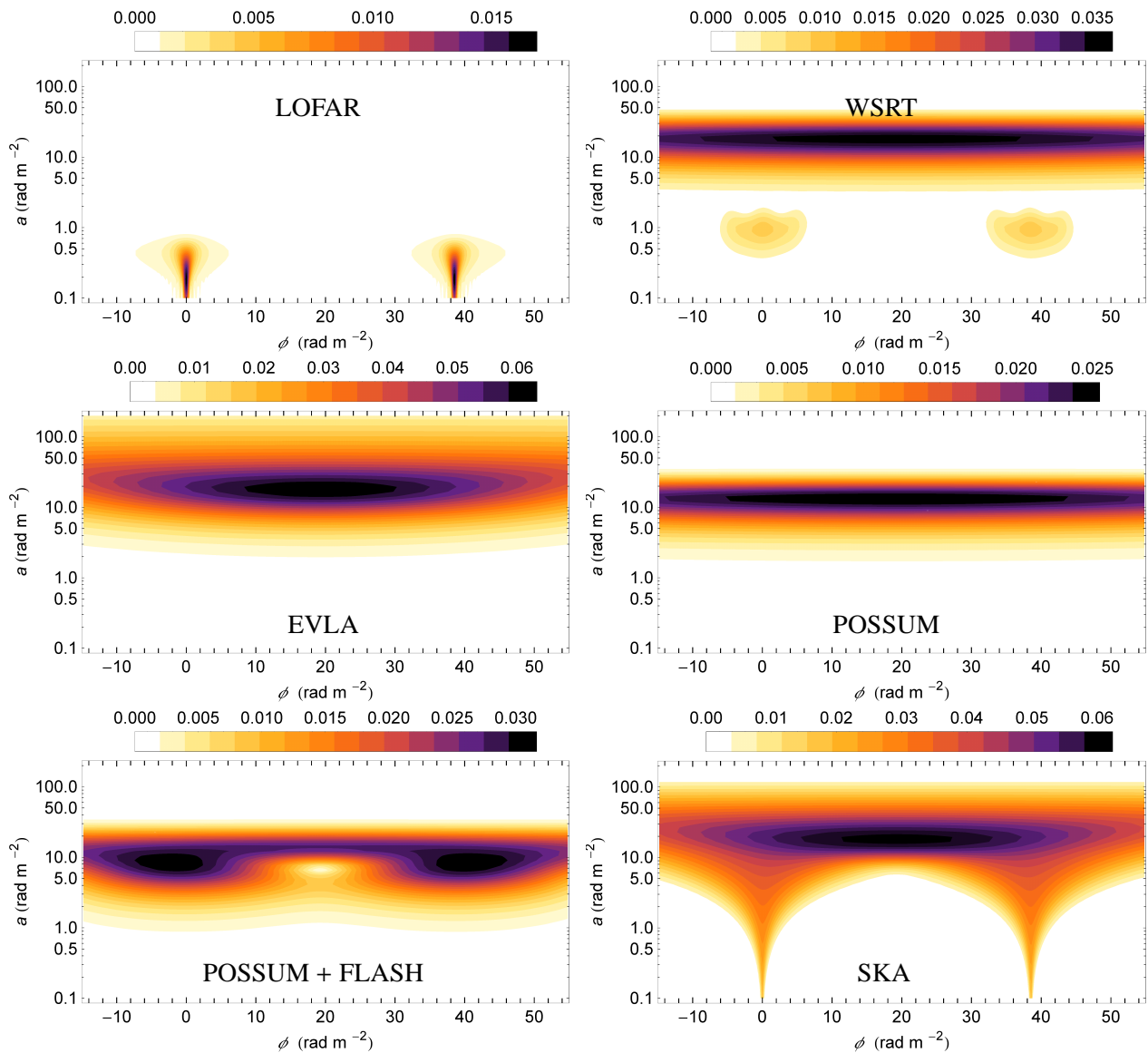}}
\put(110,390){{\large LOFAR}} \put(345,390){{\large WSRT}}
\put(115,175){{\large EVLA}} \put(340,175){{\large POSSUM}}
\put(80,30){{\large POSSUM + FLASH}} \put(350,30){{\large SKA}}
\end{picture}
\caption{Wavelet planes $w_F(a,\phi)$ for the magnetic field with
Gaussian profile and the same scale-heights of thermal and CRE
electrons (model 2). Panels are the same as in
Fig.~\protect{\ref{E1fig}}.} \label{E2fig}
\end{figure*}



\begin{figure*}
\begin{picture}(240,500)(0,0)
\put(0,000){\includegraphics[width=0.9\textwidth]{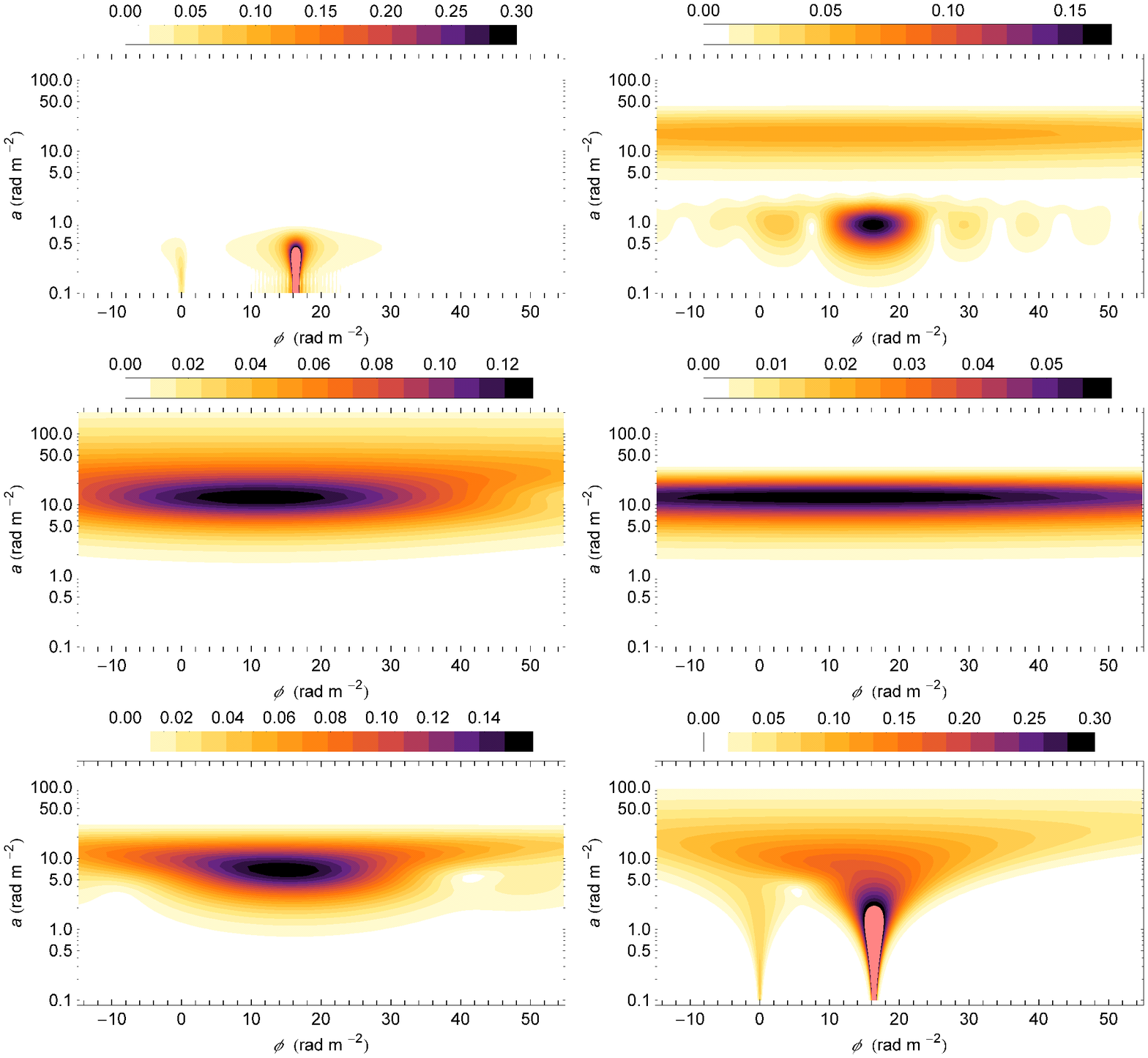}}
\put(110,390){{\large LOFAR}} \put(345,390){{\large WSRT}}
\put(115,175){{\large EVLA}} \put(340,175){{\large POSSUM}}
\put(80,30){{\large POSSUM + FLASH}} \put(360,30){{\large SKA}}
\end{picture}
\caption{Wavelet plane for the regular field with one reversal.
Panels are as in Fig.~\protect{\ref{E1fig}}. Note that the disk at
large scales is much weaker than the horn on small scales and, in
the case of the SKA in the last panel, could not be plotted within
the range of the color bar.} \label{E3fig}
\end{figure*}


- Parkes 64\,m single dish (Australia), where a southern sky survey
(S-PASS) has been performed in the frequency range 2180--2420\,MHz
\citep{carretti10}. The southern part of the GMIMS survey at
300--900\,MHz is planned.

- Effelsberg 100\,m single dish (Germany), at which a northern sky
polarization survey (GMIMS) is planned in the frequency range
300--900\,MHz.

- Arecibo 305\,m single dish (USA), where a deep spectropolarimetric
survey (GALFACTS) of the sky visible from Arecibo is ongoing in the
frequency range 1225--1525\,MHz \citep{george11}.

- Expanded VLA synthesis telescope (EVLA, USA), recently renamed as
Jansky VLA, whose new correlator (WIDAR) allows measurements in the
frequency range 1--12\,GHz (0.025--0.30\,m) with continuous coverage
at four frequency settings (L, S, C, and X-band). Polarization data
from the frequency bands at even higher frequencies are needed only
for sources with $| \Delta \phi | > 5000$\,rad/m$^2$. The first
polarization observations of the dwarf irregular galaxy IC\,10 in C
band (4.5--7.8\,GHz) were still limited in resolution in Faraday
space \citep{heesen11}. The full wavelength coverage available with
the EVLA still needs to be investigated.

- Australia Telescope Compact Array (ATCA), whose new correlator
(CABB) allows measurements in the frequency range 1.1--10\,GHz
(0.03--0.27\,m) with an almost continuous coverage at two frequency
settings (L+S and C+X band). Complex Faraday spectra were observed
in active galactic nuclei in the L+S bands \citep{sullivan12}.

- Australian Square Kilometre Array Pathfinder (ASKAP), which is a
synthesis telescope under construction in Western Australia
\citep{johnston08}. The all-sky polarization survey POSSUM in the
frequency range 1130--1430\,MHz (or 0.21--0.265\,m) is planned in
2013, together with surveys in HI and continuum. A combination of
these data with other data from the all-sky transient survey FLASH
in the frequency range 700--1000\,MHz is under discussion; this
would extend the frequency coverage to 700--1430\,MHz (or
0.21--0.428\,m) and improve the resolution in Faraday space by a
factor of about five compared to data from the POSSUM survey alone.

- Square Kilometre Array (SKA): Construction should start in 2016 in
South Africa, with extensions into central Africa, and in Australia.
The frequency coverage of the receiving systems under investigation
for ``phase 1'' (which will be available from about 2018) is about
70\,MHz--3\,GHz (or 0.1--4.3\,m) with continuous coverage, to be
extended in ``phase 2'' to 10\,GHz, opening excellent possibilities
for RM synthesis. Measuring a dense grid of RM measurements is one
of the key science projects for the SKA
\citep{gaensler04,beck10,beck11}.

Table~\ref{Tab1} summarizes the properties of the present-day and
future radio telescopes. The highest resolution in Faraday space
(largest $\Delta \lambda^2$) is achieved by LOFAR and the SKA, while
the ATCA, ELVA, and the SKA provide the largest range of scales in
Faraday space (largest $(\lambda_{\rm max}/\lambda_{\rm min})^2$,
see Sect.~\ref{secwavelet}).

The following combinations of data from synthesis telescopes yield
an excellent frequency coverage and high angular resolution, and are
considered in this paper:

- LOFAR + EVLA data of high angular resolution over a huge frequency
range (110--240\,MHz + 1--12\,GHz or 0.025--0.30\,m + 1.25--2.7\,m),
but the frequency gap hampers detection of certain structures in
Faraday space.

- LOFAR + WSRT/GMRT + EVLA data help to partly fill the gap in the
data from LOFAR + EVLA. Faraday structures can be detected on a
large range of scales, similar to the planned capability of the SKA.

Data from ATCA/ASKAP and EVLA could also be combined, but this makes
little sense because the frequency coverages are similar. Data from
ATCA/ASKAP and LOFAR are hard to combine because they cover
different parts of the sky.

The Global Magneto-Ionic Medium Survey (GMIMS) planned with several
single-dish telescopes \citep{wolleben10b} will provide a similarly
wide frequency coverage as WSRT or GMRT, but with lower angular
resolution.

\begin{figure*}
\begin{picture}(240,500)(0,0)
\put(0,000){\includegraphics[width=0.9\textwidth]{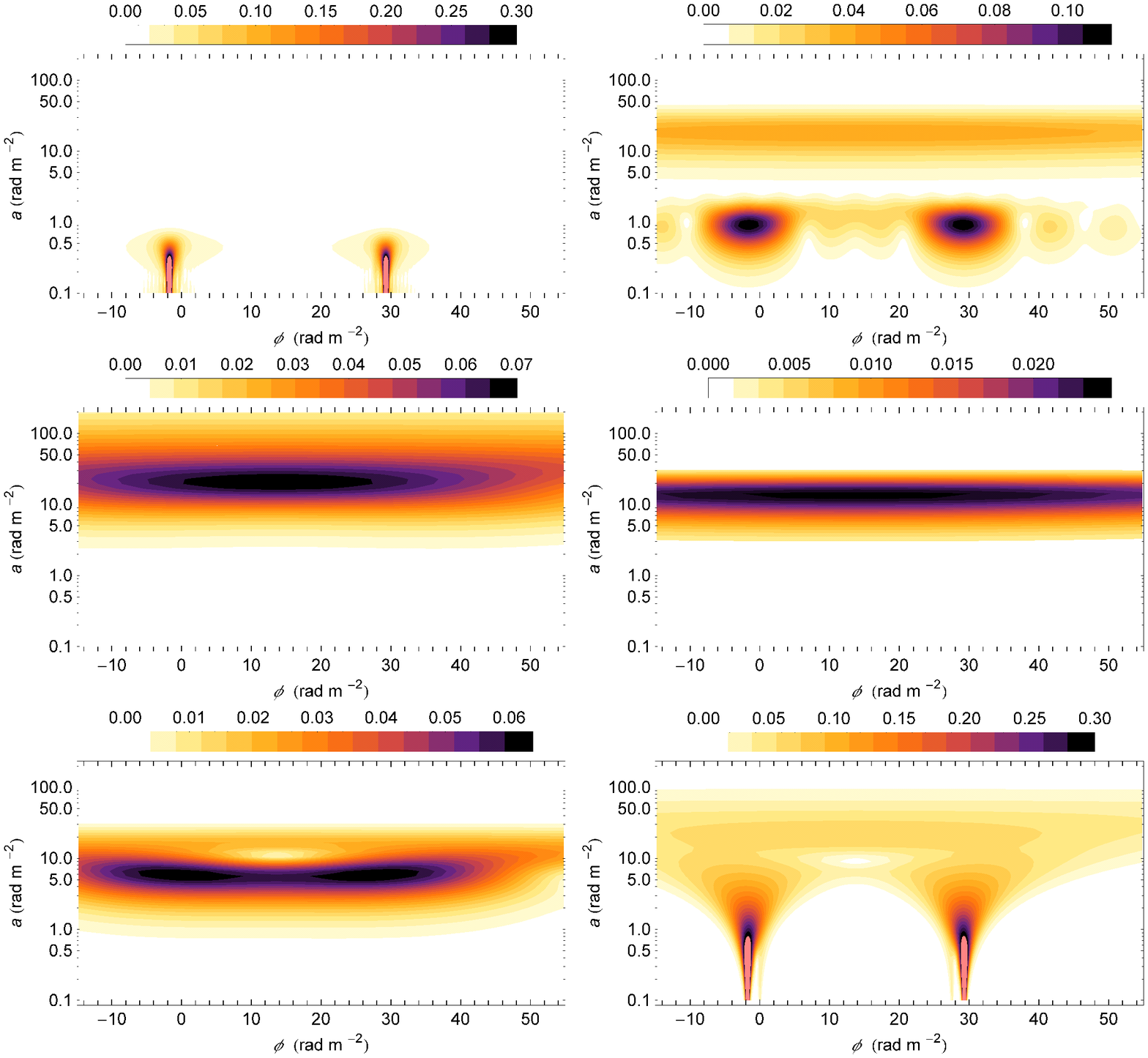}}
\put(110,390){{\large LOFAR}} \put(345,390){{\large WSRT}}
\put(115,175){{\large EVLA}} \put(340,175){{\large POSSUM}}
\put(80,30){{\large POSSUM + FLASH}} \put(350,30){{\large SKA}}
\end{picture}
\caption{Wavelet plane for the regular field with two reversals.
Panels are as in Fig.~\protect{\ref{E1fig}}.} \label{E4fig}
\end{figure*}


\section{Wavelet-based RM synthesis}
\label{secwavelet}

The wavelet transform of the Faraday spectrum $F(\phi)$ is given by
\begin{equation}
\label{wF_d}
w_F(a,\phi) = {{1}\over {a}} \int\limits_{ - \infty }^\infty
{F(\phi')\psi ^\ast \left( {\frac{\phi' - \phi}{a}} \right)\mathrm{d}\phi'} ,
\end{equation}
where $\psi(\phi)$ is the analyzing wavelet, $a$ defines the scale,
and $\phi$ is the Faraday depth of the wavelet center. We use as the
analyzing wavelet the real-value ``Mexican hat'' $\psi (\phi') =
(1-\phi'^2) \exp(-\phi'^2/2)$. The coefficient $w_F$ gives the
contribution of the corresponding structure of scale $a$ located at
position $\phi$ to the function $F$.
Wavelet-based RM synthesis was introduced by \cite{frick10} and
directly calculates the coefficients $w_F$ from the complex-valued
polarized intensities $P(\lambda^2)$. The algorithm is described in
detail by \cite{frick11} and allows us to combine the RM synthesis
procedure and the wavelet filtering. With the additional assumption
of symmetry of the radio sources, one can recover more accurately
the Faraday spectrum $F(\phi)$, particularly its complex part
yielding the intrinsic polarization angles.

We take advantage of the wavelet coefficient distribution as a
multi-scale representation of the signal. The scale analysis is
worth performing when the range of recognizable scales in Faraday
space is sufficiently wide, in other words, the ratio of maximum to
minimum wavelengths of the observations is considerably larger than
unity
\[
\frac{\Delta\lambda^2}{\lambda^2_{\rm min}}=\frac{\lambda^2_{\rm
max}}{\lambda^2_{\rm min}}-1 \gg 1 .
\]
Table~\ref{Tab1} shows that the EVLA, the ATCA, the SKA, and a
combination of LOFAR, WSRT/GMRT, and EVLA fulfil this condition. The
SKA will provide the largest value of $(\lambda_{\rm
max}/\lambda_{\rm min})^2$ in its phase 1, which will be further
increased in phase 2.

We note that a large value of $(\lambda_{\rm max}/\lambda_{\rm
min})^2$ is also needed to obtain a small error $\Delta \psi_0$ of
the intrinsic polarization angle $\psi_0$
\begin{equation}
\label{psi_0} \Delta \psi_0 = \Delta \phi \, \lambda_{\rm min}^2
\simeq \sqrt{3} \, / \, ( \,\frac{S}{N} \, (\,\frac{\lambda^2_{\rm
max}}{\lambda^2_{\rm min}}-1)\,),
\end{equation}
where $S/N$ is the signal-to-noise ratio of a ``source'' at Faraday
depth $\phi$.


\begin{figure*}
\begin{picture}(240,330)(0,0)
\put(0,000){\includegraphics[width=0.9\textwidth]{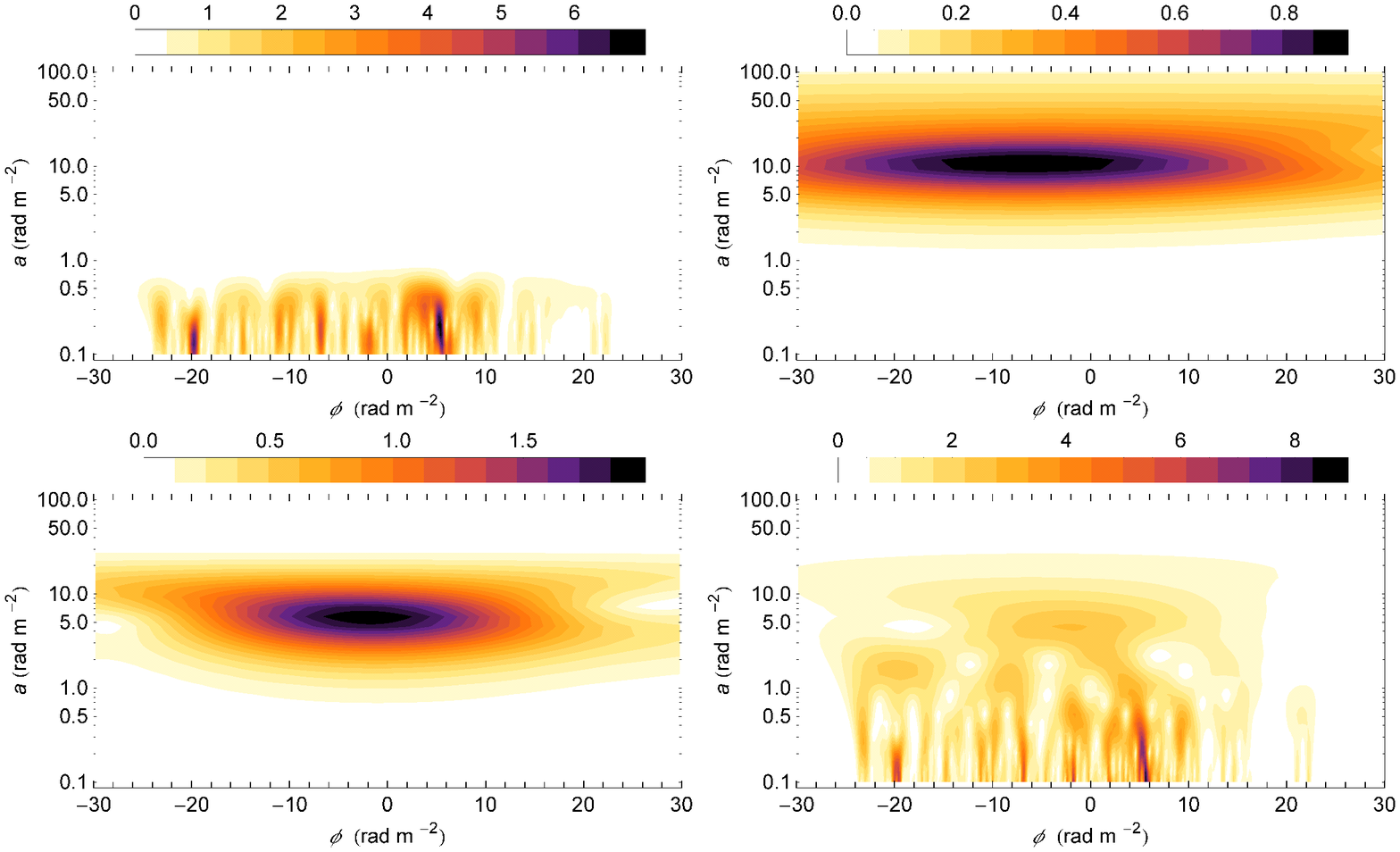}}
\put(110,245){{\large LOFAR}} \put(345,245){{\large WSRT}}
\put(115,105){{\large EVLA}}  \put(350,105){{\large SKA}}
\end{picture}
\caption{Wavelet plane for the turbulent field. Panels are as in
Fig.~\protect{\ref{E1fig}}.} \label{Ec1fig}
\end{figure*}


\section{Models}

We first consider a model region hosting a regular magnetic field
with different profiles along the line of sight, as shown in
Fig.~\ref{Bfig}. The maximum regular field strength of $\approx
2~\mu$G is typical of the Milky Way and nearby galaxies. The modeled
region also contains thermal electrons (responsible for Faraday
rotation) and cosmic-ray electrons (CRE, responsible for synchrotron
emission). Thus, the investigated region emits polarized synchrotron
emission and rotates its polarization angles according to the
Faraday effect.

\begin{figure*}
\begin{picture}(240,500)(0,0)
\put(0,000){\includegraphics[width=0.9\textwidth]{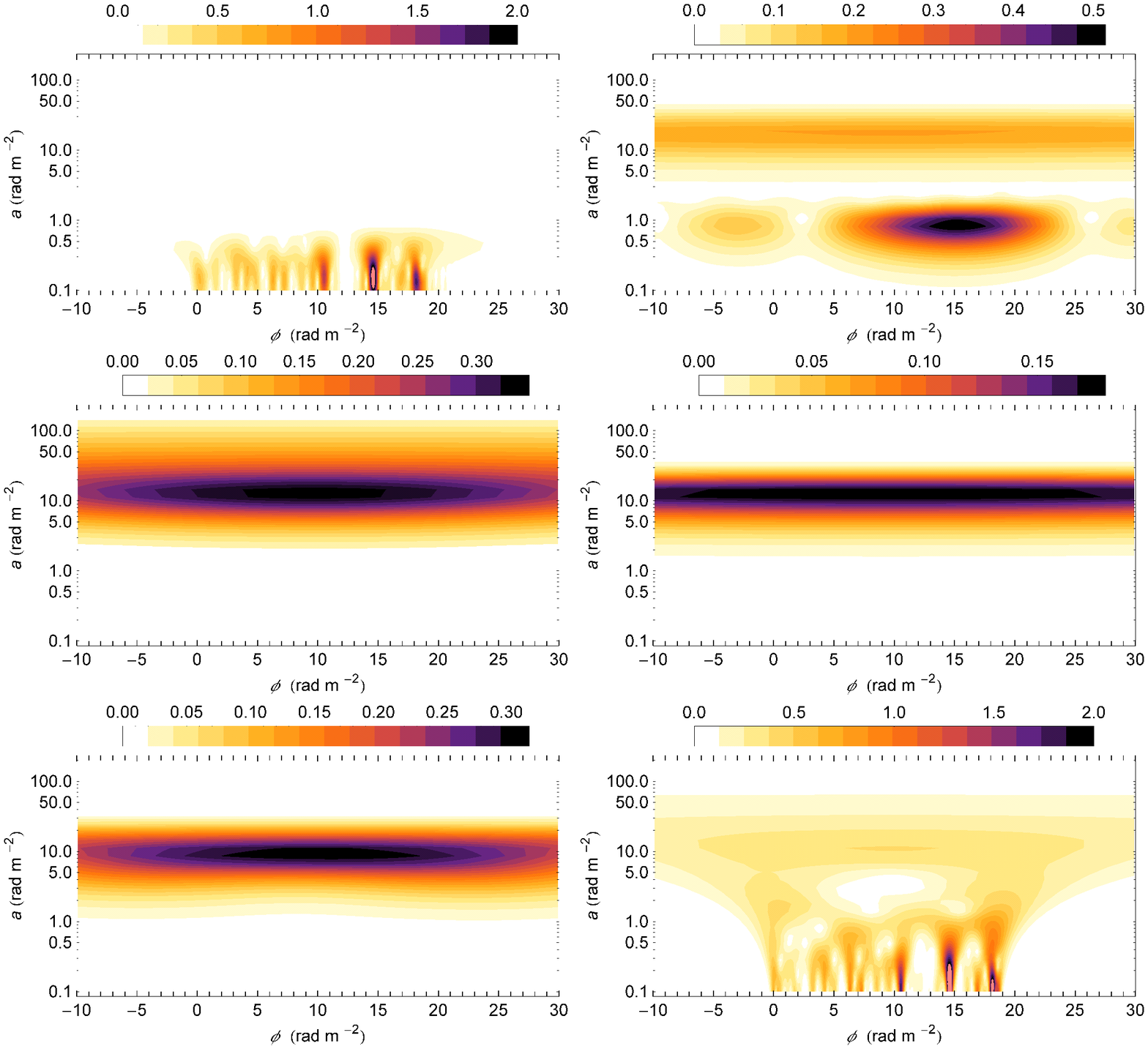}}
\put(110,390){{\large LOFAR}} \put(345,390){{\large WSRT}}
\put(115,175){{\large EVLA}} \put(340,175){{\large POSSUM}}
\put(80,30){{\large POSSUM + FLASH}} \put(410,30){{\large SKA}}
\end{picture}
\caption{Wavelet plane for the turbulent field superimposed on the
regular field. Panels are as in Fig.~\protect{\ref{E1fig}}.}
\label{Et1fig}
\end{figure*}


The distributions of relativistic electrons $n_{\rm c}$ and thermal
electrons $n_{\rm th}$ are assumed to have Gaussian profiles in the
radial direction within the plane, or with height perpendicular to
the plane
\begin{eqnarray}\label{dise}
    n_{\rm c}(x)&= &C \, {\rm exp}[-(x-x_0)^2/ h_{\rm c}^2],\\
    n_{\rm th}(x)&= &0.03\,{\rm cm}^{-3}\, {\rm exp}[-(x-x_0)^2/ h_{\rm th}^2] .
    \nonumber
\end{eqnarray}
Here $C$ is the maximum CRE number density, which is normalized to 1
in our models. Our models are applicable to galaxies observed at any
inclination angle. If the scale-radius is, for example, ten times as
large as the scale-height, the scale-height dominates the
distribution for inclinations smaller than 84\degr, which is
otherwise dominated by the scale-radius.

We assume that the Gaussian scale-height (respectively the
scale-radius) of the thermal electrons $h_{\rm th}$ is half that of
the scale-height (respectively scale-radius) of the regular magnetic
field $h_{\rm B}$. The synchrotron scale-height $h_{\rm syn}$ is
derived according to
\[
(1/h_{\rm syn})^{2} = (1/h_{\rm c})^{2} + (2/h_{\rm B})^{2} .
\]
For the CRE scale-height (resp. scale-radius) $h_{\rm c}$, four
models are considered (Table~2):

(1) $h_{\rm c} = \sqrt 2 \, h_{\rm th}  = h_{\rm B}/ \sqrt 2$. This
is the standard case expected for equipartition between the energy
densities of cosmic rays and the regular magnetic field. The region
of CRE is more extended than that of thermal electrons. The
resulting $h_{\rm syn}$ is the same as that of the thermal
electrons.

(2) $h_{\rm c} = h_{\rm th} = h_{\rm B}/2$. The regions of CRE and
thermal electrons have identical extents and $h_{\rm syn} =
\sqrt{2/3} \, h_{\rm th}$.

(3) $h_{\rm c} = h_{\rm th}/ \sqrt 2 = h_{\rm B}/ (2 \sqrt 2)$. The
region of CRE is less extended than that of thermal electrons and
$h_{\rm syn} = \sqrt{2/5} \, h_{\rm th}$. This case is expected to
occur at high frequencies and/or in regions with strong total
magnetic fields where the energy losses of CRE are high and hence
their propagation lengths are small.

(4) $h_{\rm c} = 2 \, h_{\rm th} = h_{\rm B}$. The region of CRE is
much more extended than that of thermal electrons and $h_{\rm syn} =
\sqrt{4/3} \, h_{\rm th}$. This case is expected at low frequencies
and/or in regions with weak total magnetic fields where CRE
lifetimes and hence their propagation lengths are large.

Changing the CRE scale-height leads to strong differences in the
shape of the Faraday spectrum. The first model results in a broad
Faraday spectrum $F$ with two strong ``horns'' at each end
(Fig.~\ref{Ffig}). These horns are the result of the two regions
with very low density of thermal electrons on the far and the near
side of the pathlength. One horn is correspondingly located at
Faraday depth $\phi =0$ (Fig.~\ref{Bfig}). These horns disappear for
the second model, in which both distributions have the same
scale-height. However, the Faraday spectrum $F$ still has sharp
edges. The third model with a narrow CRE distribution has neither
horns nor sharp edges. On the other hand, the extended CRE
distribution in model 4 leads to very strong horns.

We conclude that the contrast between the amplitudes of the horns
and of the broad part of the Faraday spectrum is controlled by the
relative distributions of cosmic-ray electrons and thermal
electrons. The larger the ratio $h_{\rm c}/h_{\rm th}$, the stronger
the horns. No horns are visible for $h_{\rm c} \le h_{\rm th}$.

In addition, we consider two cases with magnetic field reversals.
The distributions of relativistic and thermal electrons are the same
as in model 1, but the magnetic field includes one or two reversals
(see Fig.~\ref{Bfig}, upper panel). Reversals along the line of
sight can be the result of an axi- or bisymmetric spiral magnetic
field structure in the disk when a galaxy is observed almost
edge-on, or of an antisymmetric field structure with a reversal of
the toroidal field component above and below the galactic plane when
a galaxy is observed almost face-on. In the Milky Way, one field
reversal is observed, which is located inside the solar radius and
extends over several kpc in azimuth \citep{vaneck11}.

The typical signature of the model with one reversal is the strong
asymmetry of the Faraday spectrum $F$, called a ``Faraday caustic''
by Bell et al. (2011), which becomes symmetric for two reversals,
with two narrow features at the edges that are much stronger than in
the basic case without reversals.

Finally, having in mind objects such as the intergalactic medium in
a galaxy cluster, we consider the case of a Kolmogorov-type {\em
turbulent field}, without a regular field (Fig.~\ref{Btfig}). The
standard deviation $\sqrt \langle B_t^2 \rangle = 5~\mu$G is typical
of the intracluster medium near the center of a radio-bright galaxy
cluster. The field strength and both thermal and cosmic-ray electron
densities are assumed to have Gaussian distributions along the line
of sight with scale-radii $h_{\rm c} = h_{\rm th} = h_{\rm
B}/\sqrt{2}$.

\begin{figure*}
\begin{picture}(240,500)(0,0)
\put(0,000){\includegraphics[width=0.9\textwidth]{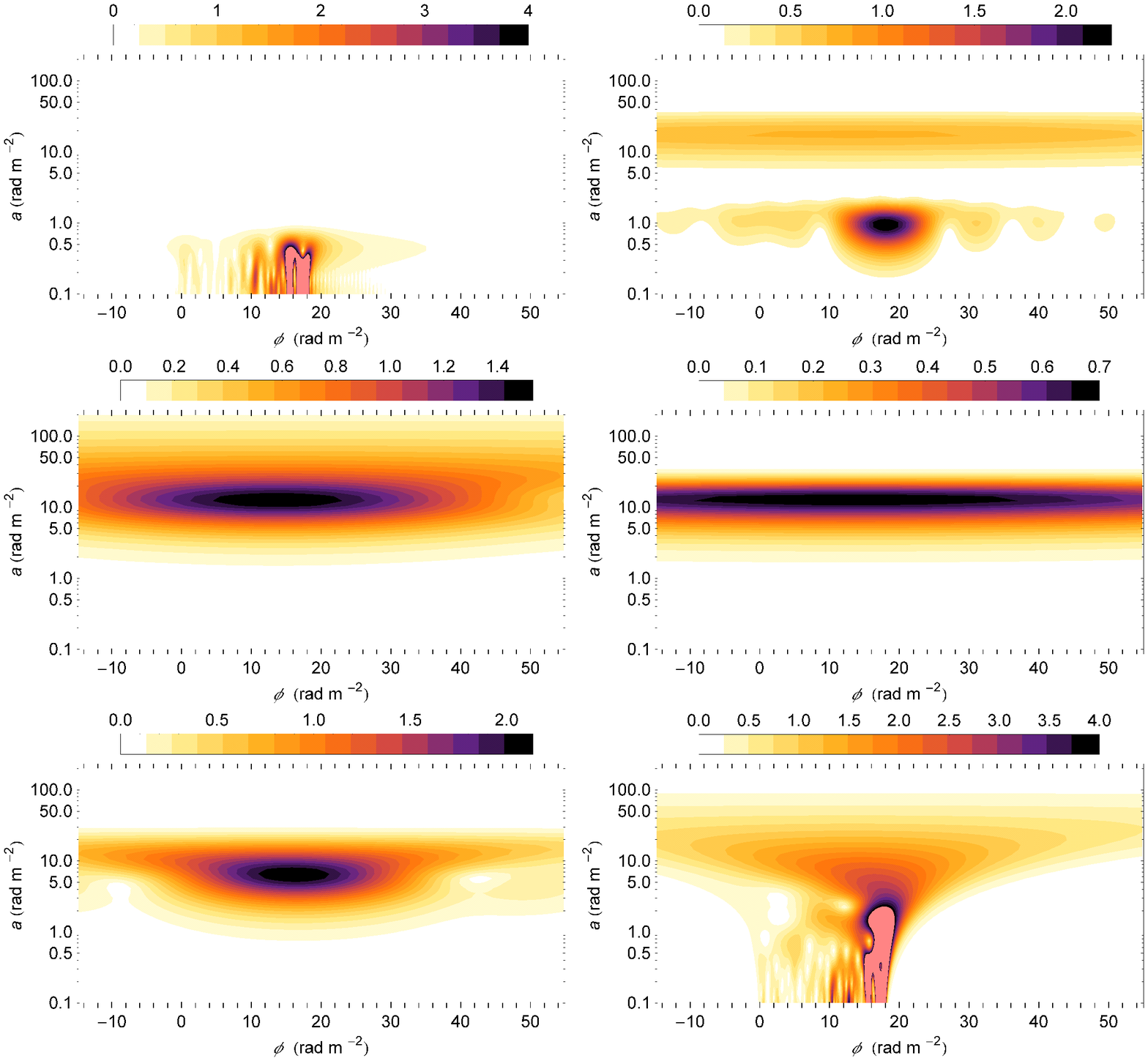}}
\put(110,390){{\large LOFAR}} \put(345,390){{\large WSRT}}
\put(115,175){{\large EVLA}} \put(340,175){{\large POSSUM}}
\put(80,30){{\large POSSUM + FLASH}} \put(380,30){{\large SKA}}
\end{picture}
\caption{Wavelet plane for the turbulent field superimposed on the
regular field with one reversal. Panels are as in
Fig.~\protect{\ref{E1fig}}.} \label{Et2fig}
\end{figure*}


\begin{figure*}
\includegraphics[width=0.95\textwidth]{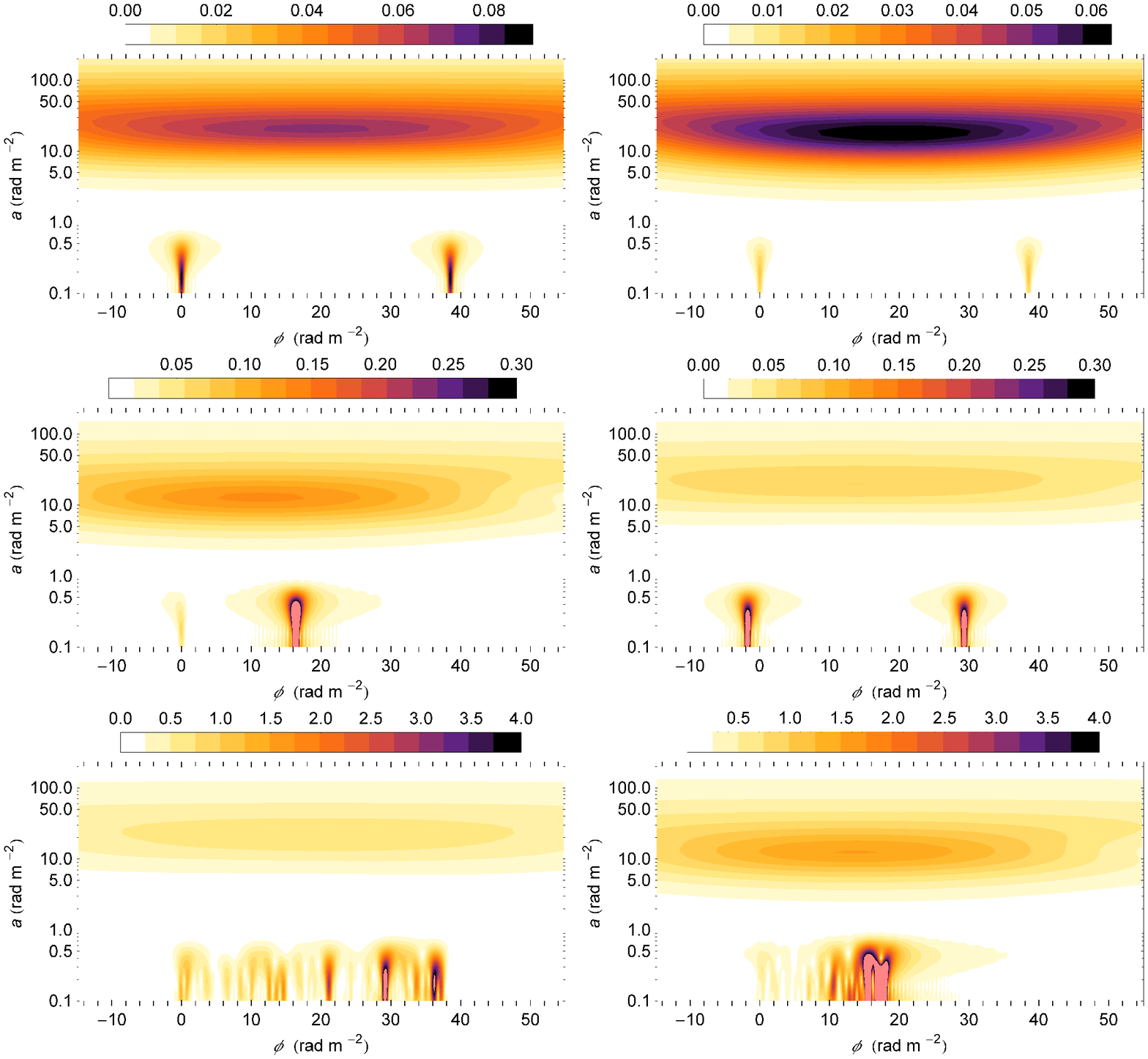}
\caption{RM synthesis applied to data from a combination of data
from LOFAR and EVLA. Panels in the left column from above: box-like
magnetic field, one reversal, and box-like field + turbulent field;
right column from above: Gaussian-like distribution, two reversals,
and regular field with one reversal + turbulent field.}
\label{C1fig}
\end{figure*}


\section{Results}

\subsection{Recognizing regular fields and reversals in magnetized regions}

\begin{figure*}
\includegraphics[width=0.95\textwidth]{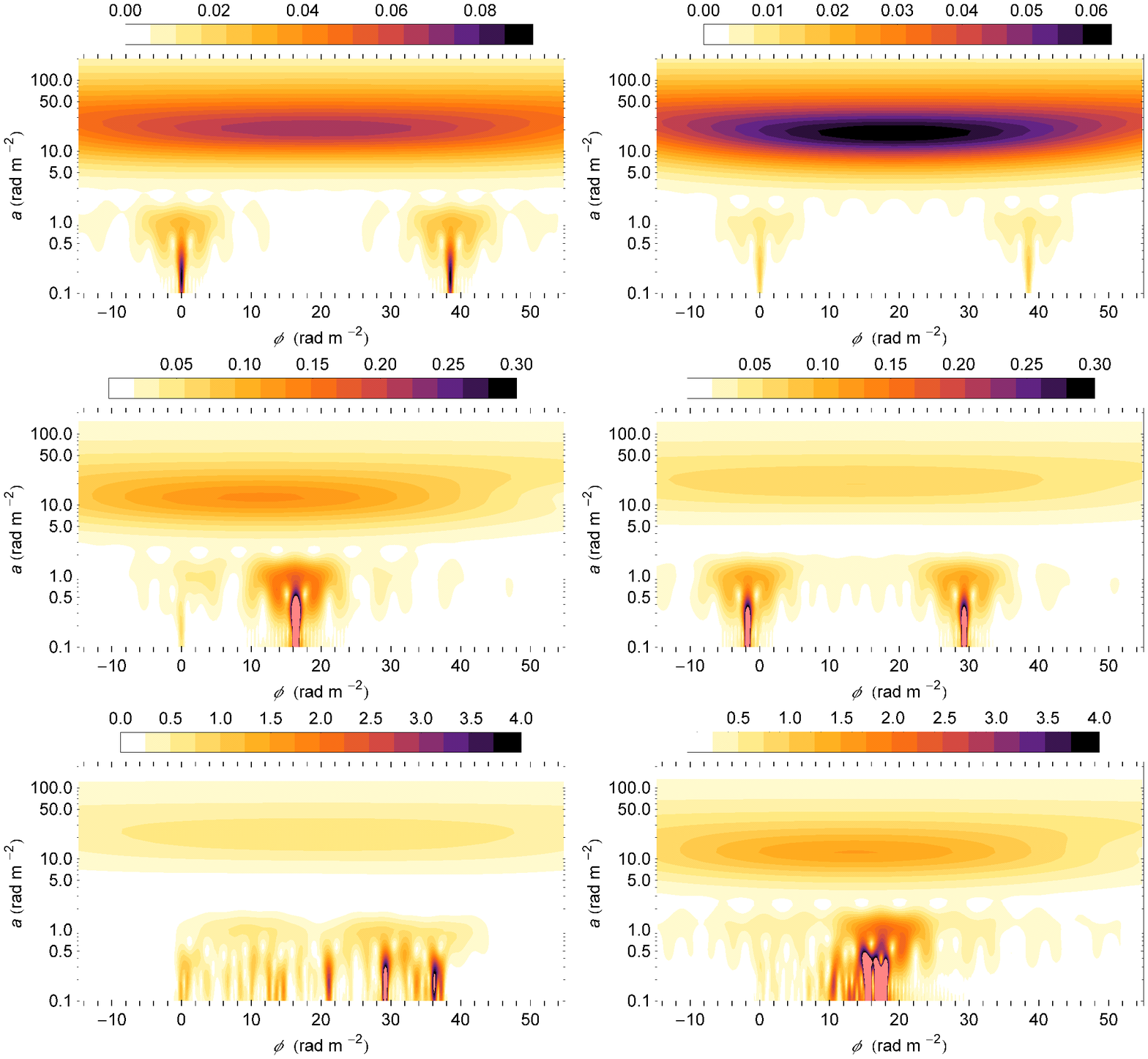}
\caption{RM synthesis for a signal from a combination of data from
LOFAR, WSRT (or GMRT), and EVLA. Panels are as in
Fig.~\protect{\ref{C1fig}}.} \label{C2fig}
\end{figure*}

As noted above, the wavelet decomposition gives a graphical
representation of a structure in the scale versus Faraday depth
plane. We present in Figures~\ref{E1fig}-\ref{C2fig} the modulus of
the wavelet decomposition of the Faraday spectrum (Eq.~\ref{wF_d}),
reconstructed from the range of $\lambda^2$ provided by various
radio telescopes, for the model examples described in the previous
section. The horizontal axis in all panels gives the Faraday depth
$\phi$ and the vertical axis shows the scale of the reconstructed
structure (also in Faraday space). The color in the wavelet plane
corresponds to the amplitude $|W_F(a,\phi)|$.

For simplicity, we assume for all wavelet transforms a constant
polarized intensity of the emitting source and a constant wavelength
response of the telescopes, to demonstrate the effects of wavelength
coverage. If the emitting source has a power-law frequency spectrum
with a negative spectral index, the lower frequencies get a higher
weight, so that the smaller scales in the wavelet plane are enhanced
with respect to the larger scales. On the other hand, Faraday
depolarization increases towards low frequencies, so that the
frequency spectrum of polarized intensity can reveal a positive
spectral index \citep{arshakian11}.

The wavelet plane for the first model (Gaussian profile of the
regular magnetic field, $h_{\rm c}=\sqrt 2 h_{\rm e}$) is shown in
Fig.~\ref{E1fig}. One horn is located at $\phi=0$, while the
location of the second horn gives the total Faraday depth of the
region. The amplitude of the two horns on small scales is larger
than that of the broad structure, which we call ``disk'' in the
following. We note that LOFAR can only recognize the two horns that
cannot be distinguished from two point sources in Faraday space.
Observing with the WSRT range picks up some of the medium-scale
structures, but the small wavelength coverage and the large gap
between the two frequency bands leads to high sidelobes that need to
be removed with ``RM clean''. With the range of short wavelengths of
ATCA, EVLA, and ASKAP (POSSUM) only the large scales are observable.
Inclusion of FLASH indicates the presence of the two horns. However,
only two broad structures are detected with ASKAP, which cannot be
distinguished from two emitting and rotating regions with similar
properties located along the line of sight. Only the SKA phase 1 can
recognize all scales and is close to a perfect ``Faraday
telescope''. Extending the frequency coverage of the SKA to 10~GHz
(0.03~m) in phase 2 would help us to observe galaxies with large $|
\Delta \phi |$.

The wavelet plane for the Gaussian profile of the regular magnetic
field and $h_{\rm c}= h_{\rm e}$ (model 2) is shown in
Fig.~\ref{E2fig}. The sharp edges of $F$ (see Fig.~\ref{Ffig})
generate horns in the wavelet plane as in the case of model 1, but
the amplitude of the horns is lower than that of the ``disk''. This
is clearly seen with the wide wavelength coverage of the SKA phase
1. Observations with ASKAP are already helpful in this case because
the scale of the transition between disk and horns lies in the range
of scales covered by POSSUM + FLASH (but not with POSSUM data
alone). The other telescopes are unable to distinguish between
models 1 and 2.

The wavelet plane for a distribution of a regular field with one
reversal is shown in Fig.~\ref{E3fig}. The dominating horn is easily
detectable with LOFAR, WSRT/GMRT, and the SKA phase 1. However,
observations at long wavelengths alone cannot distinguish this
structure from a single point source in Faraday space. Other than
for a point source, the structure of the horn is asymmetric, which
can be recognized by the inclusion of data at higher frequencies,
such as with the SKA or by combining POSSUM + FLASH. On the other
hand, the response from high frequencies alone (ATCA, EVLA, and
POSSUM) is a somewhat asymmetric disk that cannot be distinguished
from e.g. asymmetric Gaussian profiles. The second horn is weaker by
a factor of ten and is hardly detectable. Observations at several
telescope pointings or complete mapping of the source is needed to
determine the structure of sources and the extent of field reversals
in the resulting data cube.

The wavelet plane for a distribution of a regular field with two
reversals, which may exist in spiral galaxies, is shown in
Fig.~\ref{E4fig}. The horn to the left is shifted to negative $\phi$
because of the negative components of the magnetic field along the
line of sight (Fig.~\ref{Bfig}). The ``disk'' at large scales is
very weak compared to the horns, even weaker than in
Fig.~\ref{E3fig}, and is hard to detect at short wavelengths. At
long wavelengths (LOFAR and WSRT/GMRT), the two horns cannot be
distinguished from two point sources (Faraday screens). The
asymmetric nature of the two horns is visible at intermediate scales
and can be recognized from the shift of the maximum at different
scales.

Double features in the Faraday spectrum were found in the central
regions of several spiral galaxies from the WSRT SINGS survey at
1300-1763\,MHz \citep{heald+09}. As the WSRT is sensitive only to
large-scale structures in Faraday space, these features cannot be
interpreted as field reversals and are probably distinct emitting
and rotating regions along the line of sight, e.g. a regular field
component associated with the nuclear region as proposed by
\citet{heald+09}. Observations at lower frequencies are needed to
achieve higher resolution in Faraday space.

\subsection{Turbulent fields}

The signatures of turbulent fields are many components on small
scales, which we call the ``Faraday forest'', can only be observed
at long wavelengths (Fig.~\ref{Ec1fig}).
In addition, owing to the tail of the continuous turbulence spectrum
on large scales, a weaker structure is visible on yet larger scales
of $a\approx5$\,rad m$^{-2}$. This structure is less extended in
$\phi$ than the ``disk'' in the model with turbulent + regular
fields (Fig.~\ref{Et1fig}). However, observations at only high
frequencies cannot distinguish the extended structures generated by
either the tail of the turbulence spectrum or a regular field.

In real observations, the emission is smoothed over the telescope
beam and distorted by instrumental noise. The ``Faraday forest'' is
most clearly visible when the spatial scale corresponding to the
beam is similar to the turbulence scale or smaller. For a larger
beam, the ``Faraday forest'' is less visible and causes
depolarization.

Instrumental noise is spread over the whole Faraday spectrum and
hence is easy to distinguish from the ``Faraday forest'', even if
the amplitude of the ``forest'' is similar to that of noise
\citep{frick11}.

The case of a Gaussian regular field superimposed on a
Kolmogorov-type turbulent field is shown in Fig.~\ref{Et1fig}. The
standard deviation $\sqrt \langle B_t^2 \rangle$ is assumed to be
three times larger than the regular field strength, which is
approximately valid for spiral galaxies. In contrast to
Fig.~\ref{Ec1fig}, the Faraday forest extends only over a limited
range in $\phi$. On intermediate scales (WSRT/GMRT), the disk
becomes asymmetric, but no clear signatures of the turbulent field
are visible.

The corresponding result for a regular field with one reversal +
turbulent field is shown in Fig.~\ref{Et2fig}. The components of the
``Faraday forest'' are stronger at one edge of their distribution,
as can easily be recognized at long wavelengths.

\subsection{Recognizing magnetic structures by combining data from several telescopes}

Finally, we combine the wavelength ranges of LOFAR and EVLA and
apply RM synthesis for our different field models
(Fig.~\ref{C1fig}). The lack of intermediate wavelengths leads to a
separation between small and large scales in Faraday space.
Box-like, Gaussian, and double-reversal models can be distinguished
from the relative amplitudes of the horns and the disk.

By combining data from LOFAR, WSRT (or GMRT), and EVLA
(Fig.~\ref{C2fig}), structures at intermediate scales can also be
recognized and yield an almost complete picture of the magnetic
field distribution, comparable to the SKA.

\section{Conclusion and discussion}

Our conclusions are as follows:

1. We have found that a reliable recognition of magnetic field
structures in either the interstellar medium of spiral galaxies or
the intracluster medium of galaxy clusters requires us to apply RM
synthesis to spectropolarimetric data cubes observed with high
angular resolution and over a wide frequency range, from about
100\,MHz to several GHz. Such a wide frequency coverage provides
high resolution in Faraday space (as determined by $\Delta
\lambda^2$), as well as recognition of a wide range of scales in
Faraday space (as determined by $(\lambda_{\rm max}/\lambda_{\rm
min})^2$).

2. The combination of data from the POSSUM and FLASH surveys (both
planned with the ASKAP telescope) would improve the recognition of
structures on intermediate scales.

3. The combination of data from the present telescopes LOFAR and
EVLA appears to provide a promising means of recognizing magnetic
structures on all scales.

4. The combination of WSRT (or GMRT) data with those from LOFAR and
EVLA would fill the gap between the LOFAR and EVLA frequency ranges
and hence can also recognize intermediate scales in Faraday space,
which is helpful for measuring magnetic structures.

5. The detection of two ``horns'' on small and intermediate scales
plus an extended ``disk'' on large scales in the wavelet plane of
galaxies indicates that the scale-height (or scale-radius) of
cosmic-ray electrons (CRE) is larger than that of thermal electrons,
as expected especially at low frequencies. The amplitude of these
``horns'' relative to that of the ``disk'' allows the determination
of the ratio of the scale-heights (or the scale-radii) of both
electron populations. To distinguish these horns from point sources
in Faraday space, detection of intermediate scales would be
required.

6. The detection of an extended ``disk'' without ``horns'' in the
wavelet plane of galaxies indicates that the scale-height (or
scale-radius) of CRE is smaller than that of thermal electrons,
owing to e.g. the strong energy losses of CRE, which are expected at
high frequencies.

7. Recognition of field reversals in spiral galaxies needs detection
of structures on small and large scales, e.g. by combining data from
LOFAR + WSRT/GMRT + EVLA, or with the SKA, which is close to a
perfect ``Faraday telescope''.

8. Turbulent fields in galaxies or the intracluster medium can be
recognized on small scales as a ``Faraday forest'' of many
components, observable with high angular resolution and at long
wavelengths, e.g. with LOFAR or the SKA.

9. The single-dish, all-sky polarization survey GMIMS will provide
excellent resolution and scale recognition in Faraday space. The low
angular resolution allows us to investigate the structure of the
magnetized medium in the Milky Way.

10. For simplicity, the effects of instrumental noise and variations
in both angular resolution and polarized intensity with frequency
have been neglected in this paper. In practice, the combination of
data from different telescopes suffers from varying angular
resolutions and, in the case of synthesis telescopes, from different
distributions of baselines (uv coverage). Furthermore, both a
non-zero spectral index of polarized intensity and the signal
averaging within the telescope beam modify the visibility of scales
in Faraday space. Understanding these effects will require more
detailed modeling.

11. Mapping of the source is required to help us recognize the
magnetic structures in the 3-D data cubes (spatial coordinates +
Faraday depth coordinate). As the next step, the wavelet analysis
will be extended to analyzing Faraday data cubes.

12. We propose to call the data cubes generated by RM synthesis
``PPF (position--position--Faraday depth) cubes'', to be analogous
to the PPV (position--position--velocity) cubes in spectroscopy.


\begin{acknowledgements}
We thank Dominic Schnitzeler for careful reading of the manuscript,
as well as Mike Bell and Shea Brown for valuable comments. This work
was supported by the DFG--RFBR grant 08-02-92881. RB acknowledges
support from the DFG Research Unit FOR\,1254.
\end{acknowledgements}

\bibliographystyle{aa}
\bibliography{rm}

\appendix
\section{}

\begin{figure*}
\includegraphics[width=0.95\textwidth]{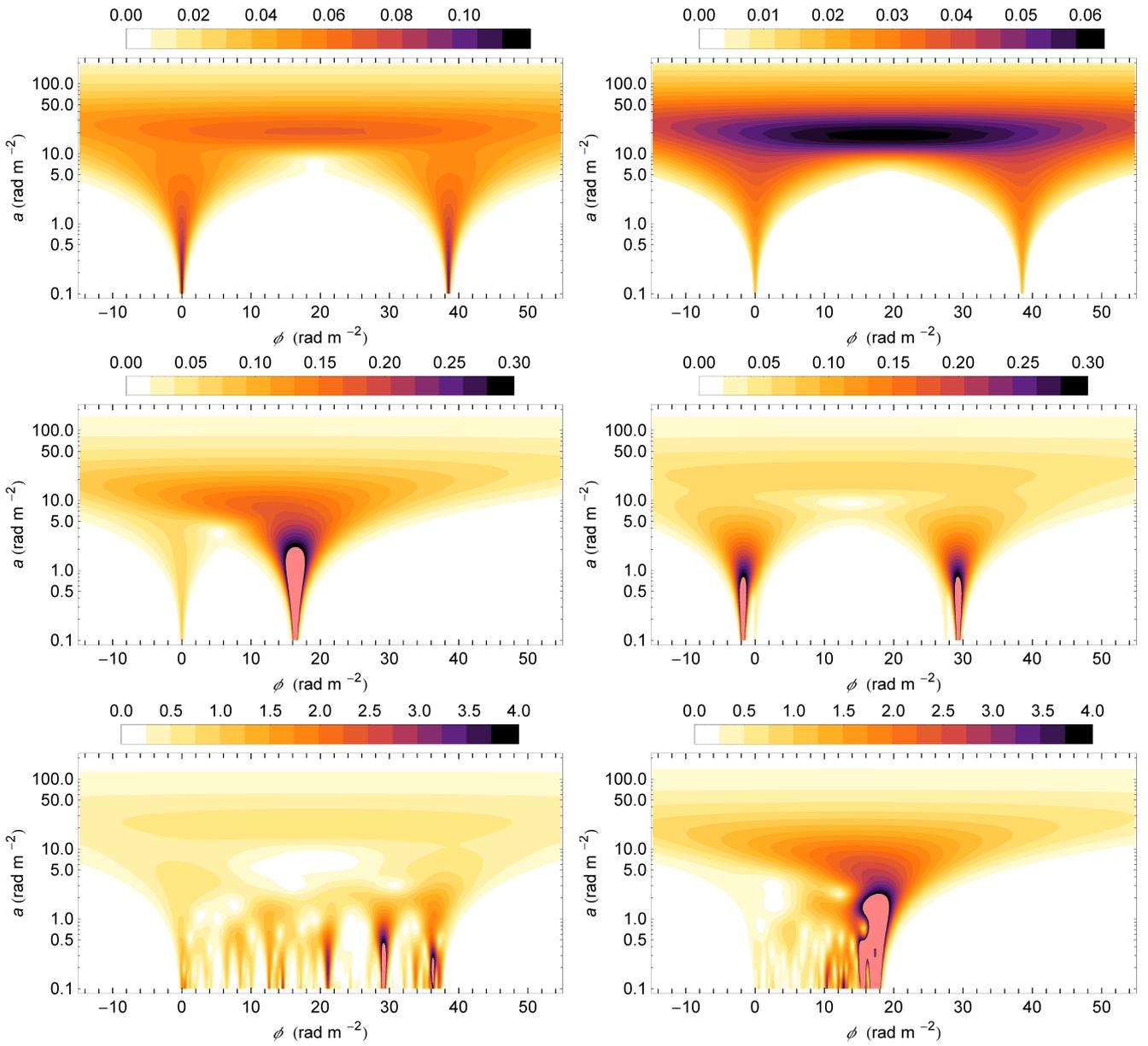}
\caption{Wavelet plane of the models considered in Sec.~5 but for
complete coverage in frequency. {\bf Top:} Regular field of model 1
(compare with Fig.~\protect{\ref{E1fig}}) and of model 2 (compare
with Fig.~\protect{\ref{E2fig}}). {\bf Middle:} Regular field with
one reversal (compare with Fig.~\protect{\ref{E3fig}}) and with two
reversals (compare with Fig.~\protect{\ref{E4fig}}). {\bf Bottom:}
Turbulent field superimposed on the regular field of model 1
(compare with Fig.~\protect{\ref{Et1fig}}) and turbulent field
superimposed on the regular field with one reversal (compare with
Fig.~\protect{\ref{Et2fig}}).} \label{Eall}
\end{figure*}

Fig.~\ref{Eall} shows the wavelet decompositions of the various
models for complete frequency sampling, as will nearly be reached by
the SKA phase 2. The result is almost the same as SKA phase 1,
except for very large scales $a>100\,{\rm rad\,m^{-2}}$.

\end{document}